\documentclass[aps,pra,twocolumn,showpacs,floatfix,superscriptaddress,a4paper]{revtex4}

\usepackage{graphicx}
\usepackage[dvipdfm]{hyperref}
\usepackage{amsmath}
\usepackage{amsfonts}
\usepackage{amssymb}
\usepackage[usenames,dvipsnames]{color}
\newcommand{\ket}[1]{|#1\rangle}
\newcommand{\bra}[1]{\langle#1|}

\newtheorem{theorem}{Theorem}
\newtheorem{problem}{Open Problem}
\newtheorem{proposition}{Proposition}

\newtheorem{definition}[theorem]{Definition}

\def\>{\rangle}
\def\<{\langle}

\begin{document}

\title{Representing probabilistic data via ontological models}

\author{Nicholas Harrigan}
\address{QOLS, Blackett Laboratory, Imperial College London, Prince Consort Road, London SW7 2BW,
UK}

\author{Terry Rudolph}
\address{QOLS, Blackett  Laboratory, Imperial College London, Prince Consort Road, London SW7 2BW,
UK}
\address{Institute for Mathematical Sciences, Imperial College London, 53 Exhibition Road,
London SW7 2BW, UK}

\author{Scott Aaronson}
\address{MIT 32-G638, Cambridge, MA 02139, USA}

\begin{abstract}
Ontological models are attempts to quantitatively describe the
results of a probabilistic theory, such as Quantum Mechanics, in a
framework exhibiting an explicit realism-based underpinning. Unlike
either the well known quasi-probability representations, or the
``r-p'' vector formalism, these models are contextual and by
definition only involve positive probability distributions (and
indicator functions). In this article we study how the ontological
model formalism can be used to describe arbitrary statistics of a
system subjected to a \textit{finite} set of preparations and
measurements. We present three models which can describe any such empirical data and then discuss how
to turn an indeterministic model into a deterministic one. This
raises the issue of how such models manifest contextuality, and we
provide an explicit example to demonstrate this.

In the second half of the paper we consider the issue of finding
ontological models with as few ontic states as possible.
\end{abstract}


\maketitle

\section{Introduction}

Entanglement plays an important role in both quantum foundations and
quantum information. Recently however, an overlap between these two
fields has emerged which goes beyond this somewhat obvious reason
for common interest. For instance, there is mounting evidence that
contextuality plays a role in any enhancement of quantum over
classical communication protocols \cite{toy_theory}. A second example
is work on the ``\emph{r-p}'' framework for analyzing the
information processing power of general (not necessarily quantum)
probabilistic theories \cite{hardy,barrett,mana,barnumbarrett} - a
body of work to which this paper is closely related. Finally we
mention work by one of us \cite{scott} showing that one particular
class of ontological models \cite{Spekkens_con,HRS} (or `hidden
variable theories') of quantum mechanics has a computational power
slightly greater than that of regular quantum computation (if access
to values of the hidden variables is granted).

As in this last example, we will  be considering ontological models.
Such models reproduce the predictions of quantum mechanics by using
(positive, normalized) probability distributions over a space of
(sometimes, though not always, hidden) variables to describe the
state of a system. Unlike quantum states in operational quantum
mechanics, these variables are presumed to correspond directly to
``properties of reality'' in some fashion - hence we term the
variables the \emph{ontic states}. Measurements in these models are
described by positive valued \emph{indicator functions} - functions
which simply determine the probability of a system in a given ontic
state yielding the particular measurement outcome associated with
that indicator function. In this way ontological models differ from
the many well known quasi-probabilty representations of quantum
mechanics. In those representations, even if the states are
represented by positive, normalized probability distributions (e.g.
the Q-function), the measurements will be represented by indicator
functions which can be negative (or greater than 1). Such negativity
is difficult to interpret if one is interested in describing
properties of a reality underpinning quantum mechanics. In fact
Spekkens has shown that such negativity arises in these
representations precisely because they are assumed non-contextual
\cite{Spekkens_private}. In particular, this assumption of
non-contextuality is made in the work on the \textit{r-p} formalism
mentioned above, and is the reason the $r$-vectors (those vectors
associated with measurements) generally have to contain negative
elements. Similar considerations apply to the work on discrete
Wigner distributions \cite{Gibbons}.

A comprehensive discussion of the ontological model framework was
undertaken in \cite{HRS}. For our purposes we will not need the
majority of the formalism developed in that paper. However, let us
mention in passing a few of the conclusions from it. Firstly, there
exist a wide variety of ontological models, both deterministic and
indeterministic, including ones which, unlike the examples given in \cite{scott},
do not provide greater computational power than that of regular
quantum computation. Secondly, for some models it is necessary to
consider not just ontic states of the system under consideration,
but also those of the preparation and measurement devices. Thirdly,
the manner in which such models exhibit contextuality is varied, but
one feature (which in some sense subsumes contextuality) that is
common to all models is a property we term \emph{deficiency}.
Loosely speaking, deficiency breaks a symmetry between preparations
and measurements in quantum mechanics. It is the property that the
set of ontic states which a system prepared in quantum state
$|\psi\>$ may \emph{actually} be in, is strictly smaller than the
set of ontic states which would reveal the measurement outcome
$|\psi\>\<\psi|$ with certainty.

In this paper, we initiate an investigation into how ontological
models can be used to reproduce quantum mechanical statistical
predictions for a discrete set of preparations of, and measurements
upon, a given system. We start in section \ref{SEC:data_tables} by
discussing a method of representing empirical data in matrix form.
In Section \ref{OFs} we then introduce a matrix factorization of
this data, equivalent to an ontological model over a finite number
of ontic states. In Sections \ref{SEC:model1}, \ref{SEC:model2} and \ref{SEC:model3}
we present three factorizations which can describe any such empirical data. We then discuss in section
\ref{DetFromIndet} how to turn an indeterministic ontological model
into a deterministic one. This raises the issue of how such models
manifest the necessary contextuality, and in Section
\ref{SEC:contextuality} we discuss this and provide an explicit
example.

In the second half of the paper, Section \ref{OC},  we consider the
issue of finding ontological models with as few ontic states as
possible - a topic with potential application to the classical
simulation of quantum systems.

\section{Data Tables \label{SEC:data_tables}}

We begin with a common \cite{hardy, mana, barrett, barnumbarrett},
although somewhat idealistic, formalization of the process of
performing an experiment resulting in probabilistic data - data for
which we then seek an explanation. The experiment consists of a set
of preparation procedures, and a set of measurement
procedures \footnote{For the majority of this paper we will ignore
`transformation procedures' (intermediate evolutions) which are also
commonly considered.}. The preparation procedures are defined in
terms of different macroscopic (and therefore distinguishable)
configurations (settings) of an apparatus, and these are labeled
$\mathcal{P}^{(i)}$, where $i=1,2,..s$ ($s$ standing for
``states''). Likewise the different measurement procedures are
macroscopically distinguishable, and can be labeled
$\mathcal{M}^{(i)}$, where $i=1,2,..m$. Each measurement procedure
has some number $d$ of outcomes, which, by padding with null
outcomes if necessary, we take to be the same for each. We can then
label the occurrence of the $k^{\text{th}}$ outcome, when the measurement
procedure is $\mathcal{M}^{(i)}$, as $\mathcal{M}^{(i)}_k$.

We imagine that a large number of experiments are performed, and the
probabilities that a given preparation yields a given measurement
outcome estimated and tabulated, e.g:
\begin{equation*}
\begin{array}{r@{\extracolsep{\fill}}l}
&\begin{array}{ccccc}
& \:\:\:\mathcal{P}^{(1)} & \mathcal{P}^{(2)} & \mathcal{P}^{(3)} & \mathcal{P}^{(4)} \\
\end{array}\\
\begin{array}{c}
\mathcal{M}^{(1)}_1\\
\mathcal{M}^{(1)}_2\\
\mathcal{M}^{(1)}_3\\
\mathcal{M}^{(2)}_1\\
\mathcal{M}^{(2)}_2\\
\mathcal{M}^{(2)}_3\\
\end{array}
&\left[\begin{array}{cccc}
 0.08 \:\:& 0.21 \:\:& ... \:\:& ... \:\:\\
 0.51 \:\:& 0.63 \:\:& ... \:\:& ... \:\:\\
 0.41 \:\:& 0.16 \:\:& ... \:\:& ... \:\:\\\hline
 0.35 \:\:& 0.72 \:\:& ... \:\:& ... \:\:\\
 0.60 \:\:& 0.21 \:\:& ... \:\:& ... \:\:\\
 0.05 \:\:& 0.06 \:\:& ... \:\:& ... \:\:\\
\end{array}\right].
\end{array}
\end{equation*}
In this example $s=4$, $m=2$ and $d=3$, so that there are $dm$ rows
in the matrix. The sum of every $d$ elements within each column must
be 1, since one of the outcomes must obtain for every measurement.
For clarity we have drawn a horizontal line underscoring the
separation between distinct measurements.

We will call such a matrix of values a data table, and generally
denote it $D$. It will often be convenient to consider $D$ a
three-dimensional array, where the particular measurement being
performed lies along the third dimension. For the above example we
can do so by defining
\begin{eqnarray*}
D^{(1)}&=&
\begin{array}{r@{\extracolsep{\fill}}l}
&\begin{array}{ccccc}
& \:\:\:\mathcal{P}^{(1)} & \mathcal{P}^{(2)} & \mathcal{P}^{(3)} & \mathcal{P}^{(4)} \\
\end{array}\\
\begin{array}{c}
\mathcal{M}^{(1)}_1\\
\mathcal{M}^{(1)}_2\\
\mathcal{M}^{(1)}_3\\
\end{array}
&\left[\begin{array}{cccc}
 0.08 \:\:& 0.21 \:\:& ... \:\:& ... \:\:\\
 0.51 \:\:& 0.63 \:\:& ... \:\:& ... \:\:\\
 0.41 \:\:& 0.16 \:\:& ... \:\:& ... \:\:\\
\end{array}\right]
\end{array},\\
D^{(2)}&=&
\begin{array}{r@{\extracolsep{\fill}}l}
&\begin{array}{ccccc}
& \:\:\:\mathcal{P}^{(1)} & \mathcal{P}^{(2)} & \mathcal{P}^{(3)} & \mathcal{P}^{(4)} \\
\end{array}\\
\begin{array}{c}
\mathcal{M}^{(2)}_1\\
\mathcal{M}^{(2)}_2\\
\mathcal{M}^{(2)}_3\\
\end{array}
&\left[\begin{array}{cccc}
0.35 \:\:& 0.72 \:\:& ... \:\:& ... \:\:\\
0.60 \:\:& 0.21 \:\:& ... \:\:& ... \:\:\\
0.05 \:\:& 0.06 \:\:& ... \:\:& ...\:\:\\
\end{array}\right]
\end{array}.
\end{eqnarray*}
More generally the data table can be taken as specified in terms of
$D^{(x)}_{i,j}$, where $x=1,2,\ldots m$, $i=1,2,\ldots d$,
$j=1,2,\ldots s$.

Now in quantum mechanics we generally associate a density matrix
$\rho_k$ with a preparation $\mathcal{P}^{(k)}$, and a POVM element
$E^{(x)}_i$ with a measurement outcome $\mathcal{M}^{(x)}_i$. Since
we will frequently be considering data tables for which the table
entries are presumed to be given by quantum mechanical expectations
- i.e. $\mathrm{Tr}(\rho_kE^{(x)}_i)$, it is often convenient to
label the rows and columns via the appropriate quantum operators. In
fact, we will almost always only be considering procedures which
correspond to pure quantum states $|\psi_k\>$ and to sets of sharp
(projective) measurements $\{\Pi_i\}$ (also known as PVM's). In this
case we have
\[D^{(x)}_{i,k}=\<\psi_k|\Pi^{(x)}_{i}|\psi_k\>,\] and it will be
convenient to adopt a more quantum notation along the lines of:
\begin{equation*}
\begin{array}{r@{\extracolsep{\fill}}l}
&\begin{array}{ccccc}
\:\:& \psi _{1} \:& \psi _{2} \:\:& \psi _{3} \:& \psi _{4} \\
\end{array}\\
\begin{array}{c}
\Pi^{(1)} _{1}\\
\Pi^{(1)} _{2}\\
\Pi^{(1)} _{3}\\
\Pi^{(2)} _{1}\\
\Pi^{(2)} _{2}\\
\Pi^{(2)} _{3}\\
\end{array}
&\left[\begin{array}{cccc}
0.08 & 0.21 & ... & ... \\
0.51 & 0.63 & ... & ... \\
0.41 & 0.16 & ... & ... \\\hline
0.35 & 0.72 & ... & ... \\
0.60 & 0.21 & ... & ... \\
0.05 & 0.06 & ... & ...%
\end{array}\right].
\end{array}
\end{equation*}%

At this stage it is pertinent to consider the following:
\begin{proposition}\label{prop1}
Every possible data table can be realized via standard quantum
expectation values of projective measurements on pure states.
\end{proposition}
At first sight this proposition may appear incorrect. Consider, for
example, the following data table:
\begin{equation*}
\begin{array}{r@{\extracolsep{\fill}}l}
&\begin{array}{ccc}
& \psi _{1} & \psi _{2} \\
\end{array}\\
\begin{array}{c}
\Pi^{(1)} _{1}\\
\Pi^{(1)} _{2}\\
\Pi^{(2)} _{1}\\
\Pi^{(2)} _{2}\\
\end{array}
&\left[\begin{array}{cc}
1 & 0\\
0 & 1\\\hline
1 & 1/2\\
0 & 1/2\\
\end{array}\right].
\end{array}
\end{equation*}%

As there are two possible outcomes for each measurement, one might
surmise that the data was generated by sharp measurements on a
qu\textbf{b}it prepared in either $|\psi_1\>$ or $|\psi_2\>$. In
that case the table is clearly impossible - the first measurement
implies the two states are orthogonal, and the second implies there
is a two outcome measurement which yields one of the states with
certainty, but which is unbiased with respect to the second state.
However, consider if the system was quantum mechanically actually a
three dimensional qu\textbf{tr}it with basis states
$|0\>,|1\>,|2\>$. In this case it is easy to verify that choosing
$|\psi_1\>=|0\>,|\psi_2\>=|1\>$, the two sets of measurements could
be $\{|0\>\<0|+|2\>\<2|,|1\>\<1|\}$ and
$\{|0\>\<0|+|+\>\<+|,|-\>\<-|\}$, where
$|\pm\>\equiv(|1\>\pm|2\>)/\sqrt{2}$. That is, the states are sharp
and the measurements are projective. The point is that there is no
way of determining from the raw data table itself any presumed
(quantum) Hilbert space dimension of the systems under
investigation. We will use the general notation $d_Q$ to denote the
Hilbert space dimension of such a quantum representation of a data
table - for this example $d_Q=3$.

Proposition~1 asserts that \emph{every} data table can be so
constructed - that is, by pure quantum states and projective
operators. That this is possible can be seen as follows. Since we
are not restricted in $d_Q$, let us imagine that every preparation
procedure (column of the data table) is represented by a state which
is orthogonal to all the other states. That is, each state lies in
its own subspace of the total Hilbert space. Clearly, if we can
choose measurements so as to reproduce any desired statistics for
one such state then, since the states are all orthogonal, we can
piece together measurement operators for the different states in a
direct sum to produce suitable measurement operators for the whole
table. Focussing on one state then, imagine the measurement under
consideration has outcomes occurring with probabilities $p_i$. All
we need to establish is that, for a given $|\psi\>$, there is a way
of choosing projection operators such that
$\<\psi|\Pi_i|\psi\>=p_i$. One way of constructing suitable
operators is to work in the basis where $|\psi\>=[1,0,0,\ldots]^T$.
Let $v$ be the row vector $[\sqrt{p_1},\sqrt{p_2},\ldots]$, and
construct the unitary matrix which has $v$ as its first row and the
remaining $d-1$ rows any orthogonal basis of the support of
$\mathbb{I}-v^Tv$. Then the columns of this unitary matrix form a
set of orthogonal states, and projectors onto these states satisfy
$\<\psi|\Pi_i|\psi\>=p_i$.

This procedure is extremely wasteful in terms of $d_Q$, the Hilbert space
dimension used to represent generic data. This raises the following:
\begin{problem}\label{hilbspacedim}
Is there an efficient procedure for finding the smallest Hilbert
space dimension required to represent an arbitrary data table in
terms of pure states and projective measurements?
\end{problem}
%
%
Although we do not have an answer to this problem, we note that being able to solve it would allow one to (approximately) solve a quantum one way communication complexity problem. Thus Open Problem 1 may benefit from studies in this field. To see the relation, note that by sending $K$ copies of systems prepared according to a procedure $\mathcal{P}^{(i)}$, a party Alice can have another party Bob estimate the probability of the $k^{th}$ outcome of some measurement $\mathcal{M}^{(j)}$ occurring. The fixed accuracy of Bob's estimation is determined by the size of $K$. Thus, if the systems realizing the preparations and measurements consist of $n$ qubits then Alice needs to send Bob $nK$ qubits for him to obtain his estimation of the desired probability. But suppose that we consider $\mathcal{P}^{(i)}$ and $\mathcal{M}^{(j)}$ to be specified by binary strings $x_i$ and $y_j$ and denote the probability that Bob is attempting to estimate by $f(x_i,y_j)$. Then requiring Bob to perform the estimation of $f(x_i,y_j)$ with Alice sending as few qubits as possible is a quantum one way communication complexity problem. Thus the efficient procedure required by Open Problem 1 would allow us to minimize $n$ and thus (given that $K$ is a constant overhead determined only by the accuracy of Bob's estimation) allow us to solve the associated quantum one way communication complexity problem.

The lesson of Proposition \ref{prop1} is that, at its most basic
level, the probabilistic structure of quantum mechanics is without
physical content. To obtain such content extra assumptions or
restrictions must be imposed. For example, assuming that
measurements on distinct systems are described by separable (tensor
products of) projection operators, the Hilbert space structure we
build to yield $D$ cannot be guaranteed to reproduce all possible
\emph{correlated} probability distributions between multiple
systems \footnote{Of course  much deeper physical content of
quantum mechanics derives from the symmetries and the Hamiltonians
by which it correctly encapsulates the dynamics of the world around
us.}.

These precautions are well known but bear repeating, because in this
paper we will be looking at representing a data table via a
different probabilistic formalism, a formalism essentially that of
classical probability theory. It is well understood that the
ontological model formalism allows for an arbitrary data table to be
represented - therefore our goal is to investigate more than simply
whether such models can reproduce empirical observations. In fact we
are primarily interested in using these models to obtain a deeper
understanding of the restrictions which occur on any proposed
realistic explanation for quantum mechanics. As such we will be
looking at the structure of such models when they are used to model
data tables corresponding to hypothetical experiments which in
quantum mechanics would be described by pure state preparations and
rank one projective measurements on a system of a known Hilbert
space dimension. Therefore we will generally have the case that
$d_Q=d$, and Open Problem \ref{hilbspacedim} will not be of concern
to us.

\section{Ontological factorization of a data table}\label{OFs}

\subsection{Defining Ontological Factorizations}\label{OFdefn}

Our primary goal is to examine factorizations of the data matrix
$D^{(x)}$, $x=1,2,\ldots m$ into the product of a $d\times\Omega$
\textit{measurement matrix} $M^{(x)}$ and an $\Omega \times s$
\textit{preparation matrix} $P$ (both positive-real valued):
\begin{equation}D^{(x)}=M^{(x)}P,\;\;\;\;\forall x=1,2,\ldots
m.\label{MPequalsD}\end{equation}

The columns of the preparation matrix $P$ are normalized probability
distributions i.e. $P$ is column-stochastic; it satisfies $0\le
P_{jk}\le 1$, and $\sum_k P_{jk}=1$ for all $j$. The interpretation
of $P$ is that  its $k^{\text{th}}$ column is  a (classical) probability
distribution, over $\Omega$ distinct \emph{ontic states}, which
corresponds to the preparation procedure $\mathcal{P}^{(k)}$ of $D$.
That is, we imagine that when a system is prepared according to
$\mathcal{P}^{(k)}$, it really is prepared in one of the $\Omega$
ontic states, and our ignorance of which particular ontic state the
system is in is represented by the probability distribution in
column $k$ of $P$.

Each row of the measurement matrix corresponds to an \emph{indicator
function} over the $\Omega$ ontic states. That is, $M^{(x)}_{ij}$ is
the probability of obtaining measurement outcome $i$, if the actual
ontic state of the system is $j$, given that measurement $x$ is
being performed.

Besides (\ref{MPequalsD}) there is one extra constraint on $M$, due
to our desire to be able to interpret an ontological factorization
as a model of the probabilistic data arising from a realistic
framework. Specifically, regardless of which ontic state a system is
in, one of the measurement outcomes \emph{must} pertain. For
instance, if we consider the $x^{\text{th}}$ measurement performed on a
system actually in ontic state $j$, then one of the $d$ outcomes
must occur. As such, we need,

\begin{equation}
\sum_{i=1}^d M^{(x)}_{ij}=1,\label{stochasticityM_constraint}
\end{equation}

for each of the $j=1\ldots\Omega$. Thus the $M^{(x)}$ are  also necessarily
column-stochastic. This is actually an important extra constraint -
without it one can reproduce quantum statistics with non-contextual
ontological models not having negative indicator functions - which has been shown not to be possible by Spekkens \cite{Spekkens_con}. (They would have a decidedly weird ontology however, as they would require a strange non-separable conspiracy between preparation and measurement devices.)

With the above constraints in mind, we can define an ontological
factorization more formally as:
\begin{definition}
A data table $D$ possesses an ontological factorization (OF) over
$\Omega$ ontic states if there exists an $\Omega \times s$,
column-stochastic matrix $P$, and  $d$  matrices $M^{(x)}$, each
$m\times \Omega$ dimensional and column-stochastic, which satisfy
$D^{(x)}=M^{(x)}P$ for all $x=1,2,\ldots d$. \label{DEF:OF}
\end{definition}

If the entries of $M$ are all 0 or 1, then the ontological
factorization is deterministic, otherwise it is indeterministic.

\subsection{Model 1: An indeterministic ontological factorization \label{SEC:model1}}

We now briefly present a trivial OF, which is easily performed for
any data table. This will be rendered
much less trivial in subsection \ref{DetFromIndet}. Specifically, in
this OF of the data table $D$ we choose \[M^{(x)}=D^{(x)} \textrm{
and } P=\mathbb{I}.\] This OF has $\Omega=s$. That is, the number of
ontic states is equal to the number of preparation procedures
(quantum states) used in constructing the data table. In terms of an
ontological model, this OF implies that preparing a system in a
quantum state $|\psi_j\>$ is equivalent to preparing the system in
a single (determined) ontic state (since each column of $P$ has only
one non-zero entry). In the terminology of \cite{HS} this model is
$\psi$-ontic (and $\psi$-complete).

Measurements in this model are represented by indeterministic
indicator functions - the model simply specifies what the
probability is that a system in a given ontic state yields each
outcome of any specified measurement.

This ontological model, when extrapolated to a ``continuum limit''
(i.e. constructed for a data table of \emph{all} measurements and
preparation procedures for a quantum system of some fixed Hilbert
space dimension), becomes the model of Beltrammati-Bugajski
\cite{beltrametti_bugajski}, which was presented as example number
$1$ in \cite{HRS}. In that model the ontological state space is
simply the complex projective space, the quantum state is
represented by a Dirac delta function distribution, and the
indicator functions over the space are indeterministic, with
probabilistic weights defined by the Born rule.

Model 1 formally encapsulates the commonly expressed viewpoint that
the quantum state \emph{is} the state of reality.

\subsection{Model 2: A deterministic ontological factorization \label{SEC:model2}}

We now turn to a deterministic OF which can be constructed for any data table. This OF has
$\Omega=d^m$ ontic states. It is convenient to index the ontic
states (that is, the rows of $P$ and columns of $M$) by an $m$-tuple
of integers of the form $(j_1,j_2,\ldots,j_m)$ where each of the
$j_x$ ranges over $1,2,\ldots,d$. We then assign probabilities over
the ontic states to create $P$ as follows:
\begin{eqnarray*}
P_{(j_1,j_2,\ldots,j_m),k}&=&\prod_{x=1}^m D^{(x)}_{j_x,k}.\\
\end{eqnarray*}
Alternatively we could specify that the $k^{\text{th}}$ column of $P$,
denoted $P_{:,k}$, is formed from the $k^{\text{th}}$ column of $D$ as
follows:\[ P_{:,k}=\bigotimes_{x=1}^m D^{(x)}_{:,k}.
\]

Consider summing over index  $j_1$:
\begin{eqnarray}\label{Peqn}
\sum_{j_1}P_{(j_1,j_2,\ldots,j_m),k}&=&\sum_{j_1}\prod_{x=1}^m D^{(x)}_{j_x,k}\nonumber\\
&=&\left(\sum_{j_1=1}^d D^{(1)}_{j_1,k}\right)\prod_{x=2}^mD^{(x)}_{j_x,k}\nonumber\\
&=&\prod_{x=2}^mD^{(x)}_{j_x,k},
\end{eqnarray}
since $D^{(1)}$ is column-stochastic. We see that in general summing
over an index simply removes all terms from the product which
contain that index. Summing over all row indices $j_x$ in this way
we confirm that $P$ is column-stochastic as required.

We now consider how to construct row $i$ of $M^{(x)}$, which
corresponds to the indicator function for $\Pi^{(x)}_i$. We assign
the value 1 to every entry in this row for which the column index
$(j_1,j_2,\ldots,j_m)$ has $j_x=i$. More formally,
\begin{equation*}
M^{(x)}_{i,(j_1,j_2,\ldots,j_m)}= \delta_{i,j_x}.
\end{equation*}
To see that this works as desired, take, for example, $i=1$, so we
are dealing with the first row of $M^{(x)}$. We compute the inner
product of this row with the matrix $P$:
\begin{eqnarray*}
&&\sum_{j_1,\ldots,j_m}M^{(x)}_{1,(j_1,j_2,\ldots,j_m)}P_{(j_1,j_2,\ldots,j_m),k}\\
&=&\sum_{j_1,\ldots,j_m}\delta_{1,j_1} P_{(j_1,j_2,\ldots,j_m),k}\\
&=&\sum_{j_2,\ldots,j_m}P_{(1,j_2,\ldots,j_m),k}\\
&=&D^{(1)}_{1,k}\sum_{j_2,\ldots,j_m}\prod_{x=2}^mD^{(x)}_{j_x,k} \\
&=& D^{(1)}_{1,k}.
\end{eqnarray*}
where we have used the observation in Eq.~(\ref{Peqn}) multiple
times. Thus we see that we reproduce the required statistics of the
data table.

An example may help illustrate the idea. Consider the following data
table with $d=3$ and $m=2$:
\begin{equation}\label{model2D}
D=\left[
\begin{array}{cc}
0 & 2/3 \\
1/3 & 1/3 \\
2/3 & 0 \\\hline
1/3 & 1/2 \\
1/3 & 1/2 \\
1/3 & 0 \\
\end{array}
\right].
\end{equation}
Following the construction above, the matrices $M$ and $P$ are given by
\begin{equation}\label{model2MP}
M=\left(
\begin{array}{ccccccccc}
1 & 1 & 1 & 0 & 0 & 0 & 0 & 0 & 0 \\
0 & 0 & 0 & 1 & 1 & 1 & 0 & 0 & 0 \\
0 & 0 & 0 & 0 & 0 & 0 & 1 & 1 & 1 \\\hline
1 & 0 & 0 & 1 & 0 & 0 & 1 & 0 & 0 \\
0 & 1 & 0& 0 & 1 & 0 & 0 & 1 & 0 \\
0 & 0 & 1 & 0 & 0 & 1 & 0 & 0 & 1 \\
\end{array}
\right), P=\left(
\begin{array}{cc}
0 & 1/3 \\
0 & 1/3 \\
0 & 0 \\
1/9 & 1/6 \\
1/9 & 1/6 \\
1/9 & 0 \\
2/9 & 0 \\
2/9 & 0 \\
2/9 & 0 \\
\end{array}
\right).
\end{equation}

There are two points to make about this model. Firstly, unlike Model 1, the columns
of the matrix $P$ from Model 2 are not orthogonal (disjoint). That is, the
probability distributions corresponding to non-orthogonal states
overlap. Such models are called $\psi$-epistemic in \cite{HS}. In
such models, knowing the ontic state of the system does not allow
one to infer with certainty what quantum state would describe the
same preparation. Unfortunately, this is only true in the case of a finite data table,
$m<\infty$. In the continuum limit of this model, the probability
distributions of non-orthogonal states become disjoint. To prove
this rigorously requires a diversion into the infinite tensor
product structures first considered by von Neumann \cite{vonN}, and
we will not do so here. We point out, however, that even though it
is $\psi$-ontic in the continuum limit, it is not, in the
terminology of \cite{HS},  $\psi$-complete. (A $\psi$-complete model
would have only one ontic state consistent with any given quantum
state.)

Finding models which are $\psi$-epistemic in the continuum limit is
not easy, some attempts were given in \cite{tr_model}, and we
believe Barrett has made some recent progress
\cite{Spekkens_private}. Because there is a continuum of quantum
states, one way to ensure an ontological model representation of all
states and measurements in a Hilbert space of dimension $d_Q$ is
$\psi$-epistemic is simply to find a model which makes use of only a
finite number of ontic states. Hardy \cite{Hardy_oeb} has shown that
this is not possible - as the number of states and measurements
increases, the number of ontic states must increase. However,
Hardy's analysis relies only on the fact that we must be able to
divide the ontic state space into disjoint sets (to represent
orthogonal measurements), and that as we go to the continuum limit
there are an infinite number of orthogonal measurements that need to
be represented in this manner \footnote{Hardy's analysis does not
actually include contextual measurements, nor does it distinguish
between sets of ontic states corresponding to $|\psi\rangle$ or
those yielding outcome $|\psi\rangle\langle\psi|$ (i.e. deficiency).
However these are easily incorporated and the conclusions remain
unchanged.}. Now the number of ways of dividing $\Omega$ ontic
states into disjoint sets grows exponentially. It would therefore seem that the
possibility remains open that the number of ontic states required for an indeterministic model might
grow only \textit{logarithmically} as the size of the data table.

But in fact this is not possible. To see this, consider the simplest case of OF's of data tables having $d=2$, which we can show must satisfy the tight bound $\Omega=\min{\left\{2^m,s\right\}}$. Clearly Models 1 and 2 set an upper bound of $\Omega\leq\min{\left\{2^m,s\right\}}$ on the required number of ontic states, by providing deterministic and indeterministic factorizations having $\Omega=s$ and $\Omega=2^m$ respectively. To give the tight bound claimed, it remains to show that for given $s$ and $m$ there always exists a data table requiring an ontological factorization having $\Omega\geq{2}^m$ if $2^m\leq{s}$ or $\Omega\geq{s}$ if $s<2^m$. We will give a general construction for a data table that lower bounds $\Omega$ and then show how it applies in the case of each of these inequalities between $s$ and $m$. First note that we can write the data tables we consider as including only the probabilities for the first outcome of each measurement (since $d=2$ the probabilities of the second outcomes are then trivially determined). Thus we can consider a data table with some parameters $\tilde{m}$ and $\tilde{s}$ having columns made up from all $\tilde{m}$-bit binary strings. For example, if $\tilde{m}=3$ and $\tilde{s}=8$ then the data table thus constructed would be,

\begin{equation}
\begin{array}{cc}
&
\begin{array}{cccccccc}
\mathcal{P}^{\left( 1\right) } & \mathcal{P}^{\left( 2\right) } & \mathcal{P}^{\left( 3\right) } &
\mathcal{P}^{\left( 4\right) } & \mathcal{P}^{\left( 5\right) } & \mathcal{P}^{\left( 6\right) } &
\mathcal{P}^{\left( 7\right) } & \mathcal{P}^{\left( 8\right) }%
\end{array}
\\
\begin{array}{c}
\mathcal{M}_{1}^{\left( 1\right) }\!\!\!\! \\
\mathcal{M}_{1}^{\left( 2\right) }\!\!\!\! \\
\mathcal{M}_{1}^{\left( 3\right) }\!\!\!\!
\end{array}
& \left[
\begin{array}{cccccccc}
~~0~~ & ~~0~~ & ~~0~~ & ~~0~~ & ~~1~~ & ~~1~~ & ~~1~~ & ~~1~~ \\\hline
~~0~~ & ~~0~~ & ~~1~~ & ~~1~~ & ~~0~~ & ~~0~~ & ~~1~~ & ~~1~~ \\\hline
~~0~~ & ~~1~~ & ~~0~~ & ~~1~~ & ~~0~~ & ~~1~~ & ~~0~~ & ~~1~~%
\end{array}%
\right].
\end{array}
\label{table_upper_bound}
\end{equation}

But an OF of any data table of this form must employ at least $2^{\tilde{m}}$ ontic states. This follows since the outcomes of measurements $\mathcal{M}^{\left(1\right)},\mathcal{M}^{\left(2\right)}$ and $\mathcal{M}^{\left(3\right)}$ are completely determined given any
of the $8$ preparations shown (i.e. the relevant entries of $D$ are all $0/1$); therefore any ontic states in the supports of
these preparations (i.e. assigned a non-zero value in the relevant column of an OF's $P$ matrix) must be deterministic (take $0/1$ values) with respect to these three measurements. Thus in the example given in (\ref{table_upper_bound}), we must use at least $8$ ontic states to deal with the $8$ distinct possibilities for assigning
outcomes to $\mathcal{M}^{\left(1\right) }$, $\mathcal{M}^{\left(2\right) }$ and $\mathcal{M}^{\left(3\right) }$. Clearly, by the same reasoning, such a data table constructed with $2^{\tilde{m}}$ preparations will require $\Omega\geq{2}^{\tilde{m}}$. Similarly, any data tables having general parameters $s$ and $m$ and containing \textit{sub}-tables of the form (\ref{table_upper_bound}) will require OFs with at least $2^{\tilde{m}}$ ontic states. Thus it only remains to determine the largest such sub-table - i.e. largest value of $\tilde{m}$ - that can fit in data tables with parameters subjected to each of the constraints $2^m\leq{s}$ and $2^m>s$. If we have $2^m\leq{s}$ then we can find a sub-table of the form (\ref{table_upper_bound}) having $\tilde{m}=m$ and $\tilde{s}=s$, giving $\Omega\geq{2^m}=\min{\{2^m,s\}}$. If however, $2^m>s$, then the largest sub-table of the form (\ref{table_upper_bound}) that we can find will have $\tilde{m}=\log_2(s)$ and $\tilde{s}=s$, yielding $\Omega\geq{s}=\min{\{2^m,s\}}$.  Thus in both cases the lower bound $\Omega\geq\min{\{2^m,s\}}$ holds. Combining this with the trivial upper bound provided by Models 1 and 2 gives the tight bound $\Omega=\min{\left\{2^m,s\right\}}$, showing that as the size of a data table is increased, the number of ontic states must grow faster than a logarithmic dependence on the parameters $s$ and $m$. The general question of how many ontic states are required to represent any particular \textit{given} data table is taken up in Section IV.


The second point to make about Model 2 is that the number of ontic states grows exponentially with the number of measurements being performed. However, this need not always be the case, as can be seen for models in which $\Omega$ depends on $m$ by invoking Caratheodory's theorem to yield a lower bound on $\Omega$ \footnote{To apply Caratheodory's theorem in lower bounding a deterministic ontological factorization's number of ontic states one must note that each ontic state in such a factorization is specified by a binary vector (the associated row of $M$), and that the entries of $D$ are generated by convex combinations of these strings (with coefficients given by the columns of $P$).}. In the next section we explicitly demonstrate such a lower bound by introducing our final model; a deterministic
ontological factorization having $\Omega =s\left( dm-1\right)$ \footnote{It may seem that the number of ontic states used by Model 3 contradicts our previous lower bound of $\Omega\geq{\left\{2^m,s\right\}}$ when $2^m\leq{s}$. But note that in this case we have $\Omega\propto{s}\geq{2^m}$.}.\

\subsection{Model 3: A deterministic ontological factorization with $\Omega=O(\text{poly}(m))$ \label{SEC:model3}}

Our final ontological factorization is tailored to employ a number of ontic states that does not increase exponentially with the number of measurements to be described. This model is best illustrated by an example of its construction. Consider the data table from (\ref{model2D}) (the method we outline is trivially adapted to an arbitrary data table). We deal with each column of $D$ separately. Suppose that we are attempting to reproduce the second column of $D$, corresponding to some preparation procedure $\mathcal{P}^{(2)}$;

\[
\left[
\begin{array}{c}
\frac{2}{3} \\
\frac{1}{3} \\
0 \\\hline
\frac{1}{2} \\
\frac{1}{2} \\
0%
\end{array}%
\right]
\]

We begin by finding the smallest entry in the chosen column, which in this case takes a value of $\tfrac{1}{3}$, and is associated with the second outcome of the first measurement. We then move this entry into another column vector, to be added to what remains of the original. We also remove the same value of $\tfrac{1}{3}$ from one of the second column entries associated with the remaining measurement - restricting our choice only to ensure that the chosen entry is greater than $\tfrac{1}{3}$. Referring to the second column of $D$ shown above, we see that there are two such entries to choose from, associated with the first and second outcomes of the second measurement and both taking the value $\tfrac{1}{2}>\tfrac{1}{3}$. We choose the entry associated with the first outcome, and thus obtain;
\[
\left[
\begin{array}{c}
\frac{2}{3} \\
\frac{1}{3} \\
0 \\\hline
\frac{1}{2} \\
\frac{1}{2} \\
0%
\end{array}%
\right] \rightarrow \frac{1}{3}\left[
\begin{array}{c}
0 \\
1 \\
0 \\\hline
1 \\
0 \\
0%
\end{array}%
\right] +\left[
\begin{array}{c}
\frac{2}{3} \\
0 \\
0 \\\hline
\frac{1}{6} \\
\frac{1}{2} \\
0%
\end{array}%
\right]
\]
The new binary column vector we have generated provides a valid set of deterministic measurement outcomes for an ontic state (weighted by $\tfrac{1}{3}$). In particular, note that by subtracting the same value from a single entry associated with each measurement we have ensured that the binary vector satisfies the stochasticity requirement of (\ref{stochasticityM_constraint}). We can repeatedly apply this procedure to the `remainder' vector $[\tfrac{2}{3},0,0,\tfrac{1}{6},\tfrac{1}{2},0]$ on the right hand side of the above expression, until only column vectors with binary entries remain. This gives;

\begin{eqnarray}
\left[
\begin{array}{c}
\frac{2}{3} \\
\frac{1}{3} \\
0 \\\hline
\frac{1}{2} \\
\frac{1}{2} \\
0%
\end{array}%
\right] &\rightarrow &\frac{1}{3}\left[
\begin{array}{c}
0 \\
1 \\
0 \\\hline
1 \\
0 \\
0%
\end{array}%
\right] +\left[
\begin{array}{c}
\frac{2}{3} \\
0 \\
0 \\\hline
\frac{1}{6} \\
\frac{1}{2} \\
0%
\end{array}%
\right] \nonumber\\
&\rightarrow &\frac{1}{3}\left[
\begin{array}{c}
0 \\
1 \\
0 \\\hline
1 \\
0 \\
0%
\end{array}%
\right] +\frac{1}{6}\left[
\begin{array}{c}
1 \\
0 \\
0 \\\hline
1 \\
0 \\
0%
\end{array}%
\right] +\left[
\begin{array}{c}
\frac{1}{2} \\
0 \\
0 \\\hline
0 \\
\frac{1}{2} \\
0%
\end{array}%
\right] \nonumber\\
&=&\frac{1}{3}\left[
\begin{array}{c}
0 \\
1 \\
0 \\\hline
1 \\
0 \\
0%
\end{array}%
\right] +\frac{1}{6}\left[
\begin{array}{c}
1 \\
0 \\
0 \\\hline
1 \\
0 \\
0%
\end{array}%
\right] +\frac{1}{2}\left[
\begin{array}{c}
1 \\
0 \\
0 \\\hline
0 \\
1 \\
0%
\end{array}%
\right] \nonumber
\end{eqnarray}

The three $0/1$ vectors generated by this procedure can be used as valid columns of an $M$ matrix, and we can take the associated probabilistic weightings to form a $P$ matrix row corresponding to the preparation $\mathcal{P}^{(2)}$. Thus by introducing three ontic states we have formed $P$ and $M$ matrices that can reproduce the the quantum statistics for all measurements from $D$ performed on the single preparation $\mathcal{P}^{(2)}$. Repeating this process to introduce a new set of ontic states for the remaining column of $P$ (corresponding to a preparation $\mathcal{P}^{(1)}$), yields the following deterministic ontological factorization,

\[
M=\left[
\begin{array}{cccccc}
0 & 0 & 0 & 0 & 1 & 1 \\
1 & 0 & 0 & 1 & 0 & 0 \\
0 & 1 & 1 & 0 & 0 & 0 \\\hline
1 & 0 & 0 & 1 & 1 & 0 \\
0 & 1 & 0 & 0 & 0 & 1 \\
0 & 0 & 1 & 0 & 0 & 0%
\end{array}%
\right] ~~,~~~~~P=\left[
\begin{array}{cc}
\frac{1}{3} & 0 \\
\frac{1}{3} & 0 \\
\frac{1}{3} & 0 \\
0 & \frac{1}{3} \\
0 & \frac{1}{6} \\
0 & \frac{1}{2}%
\end{array}%
\right]
\]

Note that since we introduce a whole new set of ontic states for each preparation, the entries of different columns of our $P$ matrix are bound to be completely disjoint. Thus, following the terminology of \cite{HS}, Model 3 can be classified as $\psi$-ontic.

The procedure outlined above will generally require introducing $dm-1$ ontic states for each preparation from $D$, which suggests $\Omega=s(dm-1)$. However, Model 3 often allows us to exceed this value, so long as the procedure is applied carefully, exploiting repeated entries in $D$ to reduce the number of ontic states we need to introduce. By this token the example given above has $\Omega=6$ - almost half of $s(dm-1)=10$.

\subsection{Deterministic ontological factorizations from indeterministic ones}\label{DetFromIndet}

In this section we will examine a process for turning the
indeterministic OF of Model 1 into a deterministic OF. First a point
of notation: from now on we will often use the two dimensional array
versions of the data table and measurement matrices: \[D=\left[
\begin{array}{c}
D^{(1)} \\\hline
\vdots \\\hline
D^{(m)} \\
\end{array}
\right],\;\;\;M=\left(
\begin{array}{c}
M^{(1)} \\\hline
\vdots \\\hline
M^{(m)} \\
\end{array}
\right),\]
so that the OF becomes simply $D=MP$.

The simplest version of the procedure we outline for recovering
determinism from indeterminism is most easily illustrated by an
example. Consider the following data table:
\[
D=\left[
\begin{array}{cc}
1/2 & 2/3 \\
1/2 & 1/3 \\\hline
3/4 & 1 \\
1/4 & 0 \\
\end{array}
\right].
\]
The OF of model 1 would have $M=D$ and \[P=\left(
\begin{array}{cc}
1 & 0 \\
0 & 1 \\
\end{array}
\right).\] That is, $\Omega=2$. The first step is to notice that
each of the probabilities in the first column of $D$ is an integer
multiple of $1/4$, while each in the second column is an integer
multiple of $1/3$. We therefore expand the supports of the
probability distributions in $P$ from single ontic states to 4 and 3
ontic states respectively:
\[P=
\left(
\begin{array}{cc}
1/4 & 0 \\
1/4 & 0 \\
1/4 & 0 \\
1/4 & 0 \\
0 & 1/3 \\
0 & 1/3 \\
0 & 1/3 \\
\end{array}
\right).
\]
Note that the probability distributions are still disjoint (in the
language of \cite{HS} the model is $\psi$-ontic but not
$\psi$-complete). We now want to choose a matrix $M$ with only 0/1
entries. If we choose the first row of $M$ (i.e. the indicator
function for the first measurement outcome) to be
\[
\left(
\begin{array}{ccccccc}
1 & 1 & 0 & 0 & 1 & 1 & 0 \\
\end{array}
\right),
\]
then we will certainly produce the first row of $D$. The key is that
the disjointness of the two columns of $P$ enables us to assign 1's
and 0's to $M$'s first row in an essentially unrestricted fashion - the first 4
elements of this row simply need to pick up enough $1/4$'s from $P$
to give the entry 1/2 for $\psi_1$, and similarly the last three
entries must pick up an appropriate number of the 1/3's to give the
entry 2/3 for $\psi_2$. Assigning the second row we need to be more
careful. One assignment which naively would seem to work, in as much
as it would reproduce the data table entries when multiplied by $P$,
is
\[
\left(
\begin{array}{ccccccc}
1 & 1 & 0 & 0 & 1 & 0 & 0 \\
\end{array}
\right).
\]
However this violates the constraint of column-stochasticity on the
$M^{(x)}$'s from Definition \ref{DEF:OF}. Such an indicator function
would imply that a system in ontic state 1 gives both the first
measurement outcome and the second outcome with certainty! The only
choice for this row once we have chosen the first row as above is
\[
\left(
\begin{array}{ccccccc}
0 & 0 & 1 & 1 & 0 & 0 & 1 \\
\end{array}
\right).
\]
Proceeding in this manner, one possibility for $M$ is
\[
\left(
\begin{array}{ccccccc}
1 & 1 & 0 & 0 & 1 & 1 & 0 \\
0 & 0 & 1 & 1 & 0 & 0 & 1 \\\hline
1 & 1 & 1 & 0 & 1 & 1 & 1 \\
0 & 0 & 0 & 1 & 0 & 0 & 0 \\
\end{array}
\right).
\]

Imagine now that the data table only contains rational fractions as
entries. This would certainly be the case if we were considering a
table which was assembled from actual experimental data. Clearly,
generalizing the above procedure will allow us to find $M$ and $P$
matrices so as to obtain a deterministic OF. This does not give us a procedure for obtaining a finite OF for \textit{any} data table however, since we may wish to consider ones constructed from quantum states
and projectors which have irrational valued overlaps. For the
general program of research into ontological models there is no
requirement of finiteness  - in particular the trick of spreading
the support of the probability distribution over extra ontic states
can readily be performed by spreading the distribution over
variables of cardinality $2^{\aleph_0}$. This yields a procedure for
transforming any indeterministic ontological model into a
deterministic one - and is essentially what Bell did
\cite{Bell_first} (see example 4 in \cite{HRS}) to provide a
counterexample to `von-Neumann's silly assumption'. In doing so one
has increased the number of ontic states required to describe the
system - the extent to which this is necessary, as opposed to merely
sufficient, is an interesting question, which to some extent we take
up again in Section \ref{OC}.

\begin{figure}[t]
\includegraphics[scale=0.7]{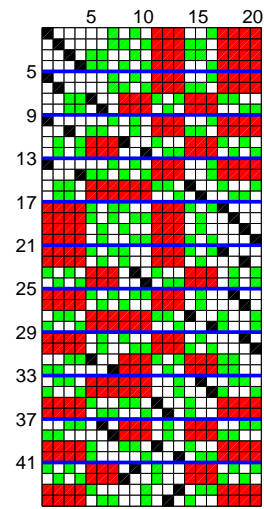}
\caption{(Color online) The data table of quantum mechanical
statistical predictions for the Kernaghan set of 11 measurements and
20 states. The values of the entries are color coded as black,
green, red and white for $1$, $1/2$, $1/4$ and $0$ respectively. The
blue lines separate different 4-outcome PVMs.}
\label{FIG:peresD}%
\end{figure}

In the last four sections we have derived a series of bounds on the number of ontic states that ontological factorizations must employ. By a simple extension of the argument given in Sec.~\ref{SEC:model2} we have that data tables with arbitrary parameters $s,d,m$ require a number of ontic states that is tightly bounded by $\Omega=\min{\{d^m,s\}}$. However, the explicit construction that we gave in Sec.~\ref{SEC:model3} showed that by letting $\Omega$ depend on all three parameters of a data table, the number of ontic states need not depend exponentially on $m$. These bounds clearly have interesting repercussions for the ability to classically simulate quantum systems efficiently. For example, our lower bound $\Omega\geq\min{\{d^m,s\}}$ shows that one cannot precisely reproduce a general data tables quantum statistics by keeping track of a number of ontic states that grows only logarithmically with the number of preparations and measurements considered. Extending and refining the bounds we have presented is an interesting direction for future research, which may benefit from existing results in the literature. For example, Aaronson has shown in \cite{aaronson_advice} that the probabilities for a set of $m$ two-outcome measurements performed on an $N$ dimensional quantum system can be approximately calculated (to some fixed accuracy) from a classical string of length $O\left(\log(N)\log\log(N)\log(m)\right)$. This shows that a data table containing $m$ such measurements and some set of preparations associated with an $N$ dimensional quantum system has an \textit{approximate} ontological factorization that employs $\Omega=2^{O\left(\log(N)\log\log(N)\log(m)\right)}=m^{O\left(\log(N)\log\log(N)\right)}$ ontic states (wherein each ontic state encodes a possible instance of the string one can construct to encode the required statistics). Thus, in cases where one is satisfied with \textit{approximately} reproducing the entries of a data table, the results in \cite{aaronson_advice} give a lower bound on $\Omega$ which depends pseudo-polynomially on $m$ and $N$, but interestingly not on $s$ (the precise number of preparations in the data table).

In Sec.~\ref{SEC:matrix_connection} we will also note how existing results concerning matrix factorization problems might prove useful in deriving more bounds on $\Omega$.

\subsection{Contextuality in ontological factorizations \label{SEC:contextuality}}

\begin{figure*}[t]
\includegraphics[scale=0.8]{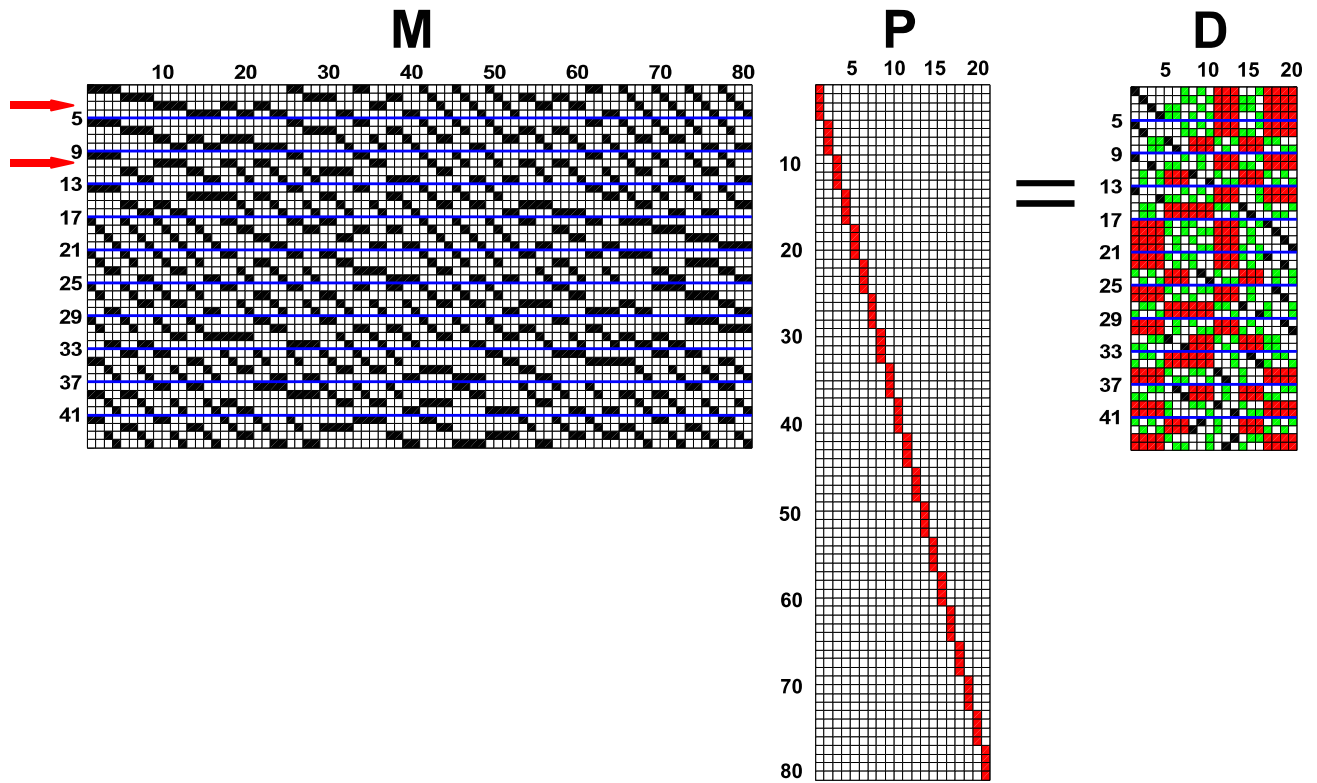}\caption{(Color online) P and M matrix for a deterministic OF of the Kernaghan set of
preparations and measurements. The black squares denote a value of
$1$, the red of $1/4$ whilst the white squares represent $0$.}
\label{FIG:peresPM}
\end{figure*}

By the famous result of Kochen and Specker \cite{Ks} deterministic
ontological models must be measurement-outcome contextual. We now
consider how measurement outcome contextuality manifests itself in
an OF of a data table. Contextuality needs to be considered when
different quantum measurements share a common projector. In such a
case a data table might look something like (with $d=3$):
\begin{equation}\label{redundanttable}
\begin{bmatrix}
& 0.08 & 1 & 0.67 & 0.09 \\
& 0.51 & 0 & ... & ... \\
& 0.41 & 0 & ... & ... \\\hline
& 0.08 & 1 & 0.67 & 0.09 \\
& 0.92 & 0 & ... & ... \\
& 0.00 & 0 & ... & ...%
\end{bmatrix}.
\end{equation}%

Rows 1 and 4 contain the same entries, and so in quantum mechanics
they could correspond to measurement outcomes represented by the
same projector. Initially one might have been tempted to simply
exclude the redundant rows. The essence of the Kochen-Specker
theorem is that doing so would prevent us being able to explain the
data table in terms of a deterministic ontological model, and so we
retain them. It is at this point that we depart from the
\textit{r-p} formalism. In the \textit{r-p} formalism there is a
unique $r$-vector for each quantum projector. As mentioned above,
this presumption of measurement outcome non-contextuality is the
reason the $r$-vectors require negative entries (or entries greater
than 1)\footnote{In fact they generically have negative entries even
when the data table being reproduced does not require a
non-contextual representation.}.

In terms of our OF's, the contextuality required by the
Kochen-Specker theorem manifests itself as follows: If two rows of
the data table are identical, then one cannot necessarily find an OF
such that the two corresponding rows of $M$ are identical. Now, if
the only constraint on the factorization was that $D=MP$ with $M$ a
$0/1$ matrix, then this would \emph{not} be true, since there would
be nothing stopping the replacement of one of those rows of $M$ by
the other. The key issue, as mentioned above, is the extra
column-stochasticity  constraint on the factorization, which forces
at least one measurement outcome to be obtained for any given ontic
state.

Let us demonstrate all this with an example. In \cite{kernaghan}
Kernaghan provided an example of a set of measurements in a 4
dimensional Hilbert space for which a Kochen-Specker obstruction
exists (see also \cite{peres}). These measurements are built from
the following 20 states in a Hilbert space of dimension 4 (the
normalization factors of $1/\sqrt{2}$ or $1/2$ are omitted, and
$\bar{1}$ denotes $-1$):
\[
\begin{array}{rllll}
|\psi_1\>=&    (1  &   0  &   0  &   0),\:\:\\
|\psi_2\>=&    (0  &   1  &   0  &   0),\:\:\\
|\psi_3\>=&    (0  &   0  &   1  &   0),\:\:\\
|\psi_4\>=&    (0  &   0  &   0  &   1),\:\:\\
|\psi_5\>=&    (0  &   0  &   1  &   1),\:\:\\
|\psi_6\>=&    (0  &   0  &   1  &  \bar{1}),\:\:\\
|\psi_7\>=&   (1  &  \bar{1}  &   0  &   0),\:\:\\
|\psi_8\>=&   (0  &   1  &   0  &   1),\:\:\\
|\psi_9\>=&   (1  &   0  &   1  &   0),\:\:\\
|\psi_{10}\>=&  ( 0  &   1  &   0  &  \bar{1}),\:\:\\
\end{array}
\begin{array}{rllll}
|\psi_{11}\>=&  ( 1  &   1  &   1  &   1),\\
|\psi_{12}\>=&  ( 1  &  \bar{1}  &  \bar{1}  &   1),\\
|\psi_{13}\>=&  ( 1  &   1  &  \bar{1}  &  \bar{1}),\\
|\psi_{14}\>=&  ( 0  &   1  &   1  &   0),\\
|\psi_{15}\>=&  ( 0  &   1  &  \bar{1}  &   0),\\
|\psi_{16}\>=&  ( 1  &   0  &   0  &  \bar{1}),\\
|\psi_{17}\>=&  (\bar{1}  &   1  &   1  &   1),\\
|\psi_{18}\>=&  ( 1  &  \bar{1}  &   1  &   1),\\
|\psi_{19}\>=&  ( 1  &   1  &  \bar{1}  &   1),\\
|\psi_{20}\>=&  ( 1  &   1  &   1  &  \bar{1}).\\  \end{array}
\]
The 11 sets of 4-outcome measurements which lead to a Kochen-Specker obstruction can be tabulated as follows:
\begin{equation}\label{perestable}
\left(
\begin{array}{rrrrrrrrrrr}
1&    1&     1&     1&    17&    17&    18&    19&    15&     6&     9\\
2&    2&     3&     4&    18&    19&    19&    20&    16&     7&     8\\
3&    5&     8&    14&    19&     9&    14&     5&    11&    11&    13\\
4&    6&    10&    15&    20&    10&    16&     7&    12&    13&    12\\
\end{array}
\right).
\end{equation}

So, for example, the third measurement $\Pi^{(3)}$ consists of the
projectors
$\{\ket{\psi_1}\bra{\psi_1},\ket{\psi_3}\bra{\psi_3},\ket{\psi_8}\bra{\psi_8},\ket{\psi_{10}}\bra{\psi_{10}}\}$.
The proof that  these projectors cannot be assigned a unique truth
value (i.e. independent of the other projectors in the measurement)
is trivial.  In each of the odd number (i.e.  11) of measurements,
only one projector can (and must) be assigned the truth value 1.
This is in direct conflict with the fact that each of the
$|\psi_i\>$'s appears an even number of times in (\ref{perestable}).

The data table for these 11 measurements and the 20 states is
depicted in a color coded fashion in Figure \ref{FIG:peresD}.
Applying the procedure of section \ref{DetFromIndet} allows us to
generate matrices $M$ and $P$ for a deterministic OF over $20\times
4=80$ ontic states. The matrices for this OF are depicted in Figure
\ref{FIG:peresPM}.

An example of the contextuality of this OF can be seen by
considering rows $3$ and $10$, which correspond to the same
measurement projector $|\psi_3\>\<\psi_3|$ (these rows are marked on
$M$ in Figure \ref{FIG:peresPM} by red arrows). Clearly each of
these indicator functions must have a 1 at all of the set of ontic
states which $|\psi_3\>$ has support over (c.f. the third column in
$P$ - that is, ontic states 10 through 13). We see, however, that
there are some ontic states (71, 75 and 79 for example) that do not
lie in the support of $|\psi_3\>$), and to which one of rows $3$ or
$10$ assigns a 1 while the other assigns a 0. Ontic states such as
71, 75 and 79 also show that there are some ontic states which are
assigned a value $1$ by the row in $M$ corresponding to
$|\psi_3\rangle\langle\psi_3|$ but not by the column in $P$
corresponding to $|\psi_3\rangle$. As discussed in \cite{HRS} this
is a general (necessary) feature of all ontological models: there
are necessarily some ontic states which pass a test $|\psi\>\<\psi|$
for a system ``being in'' a state $|\psi\>$, which can never be
prepared when the preparation procedure specifies the quantum state
to be $|\psi\>$. This asymmetry of deterministic ontological models
between preparations and measurements, which is not present in
quantum mechanics, was termed \emph{deficiency} in \cite{HRS}. The
terminology comes from the fact that the support of the probability
distribution corresponding to the quantum state is strictly less
than the support of the indicator function corresponding to a
projection onto that state, as we have seen. We term
\emph{unfaithful} those ontic states which do not lie in the support
of the probability distribution, but which \textit{do} pass the test
(since for some measurements they will ``choose'' to give a
different outcome). Note that if an ontological model is therefore
to be able to describe preparing the state $|\psi\>$ by performing a
measurement of $|\psi\>\<\psi|$, it is going to be the case that a
system in an unfaithful ontic state must be disturbed by the
measurement in order to end up in one of the faithful states.



\section{Ontological Compression}\label{OC}

So far we have investigated OF's without particular regard to how
large the number of ontic states $\Omega$ is required to be. In this
section we consider the problem of trying to either find OF's with
$\Omega$ as small as possible, or to reduce the value of $\Omega$
for a given OF - a process we call ``ontological compression''. Such a reduction in ontic states clearly has repercussions for the ability to efficiently simulate quantum systems classically - a point we elaborate on in Sec.~\ref{SEC:class_simulation}. But, as we shall see, our ontological compression schemes also provide routes towards constructing $\psi$-epistemic models in which we can truly view the quantum description of a system as a state of knowledge.

Models 1 and 2 gave OF's which require $\Omega=s$ and
$\Omega=d^m$ respectively - these values set upper bounds on
$\Omega$ for  indeterministic and deterministic OF's respectively. But since deterministic ontological factorizations can be seen as a special
case of indeterministic ones then Model 2 also provides an upper bound of $\Omega =d^{m}$ for \textit{indeterministic} factorizations. There are therefore upper bounds on $\Omega$ for both deterministic and indeterministic models which depend only on a data tables measurements, and upper bounds for indeterministic models which depend on only $s$. One may therefore wonder whether an upper bound can be found for deterministic models which
depends only on $s$. But this is not possible; a deterministic ontological factorization always
requires at least $dm+1$ ontic states.

To see why, note that each ontic state, $\lambda_i$, from a deterministic model is associated with a length $dm$ binary string, given by the $i^{\text{th}}$ column of the models $M$ matrix. Thus the ontic states of any deterministic ontological factorization can be thought of as column vectors, $\vec{\lambda}_{i}$, defining vertices on a unit hypercube of dimension $dm$; $\mathfrak{C}^{dm}\subset\left[ 0,1\right]^{dm}$. Now consider how such a model reproduces the $k^{\text{th}}$ column of the data table $D$ (corresponding to the measurement statistics for some preparation $\mathcal{P}^{\left(k\right)}$) in terms of these $\vec{\lambda}_i$ column vectors. Writing the relevant column of $D$ as a vector $\vec{p}^{(k)}\in\left[0,1\right]^{dm}$ containing the outcome probabilities for each measurement, its entries are reproduced by a convex combination of the $\vec{\lambda}_{i}$,
\begin{equation}
\sum_{i=1}^{\Omega}q^{(k)}_{i}\vec{\lambda}_{i}=\vec{p}^{(k)}.  \label{convex_decomp}
\end{equation}
Where we use $q^{(k)}_i$ to denote the $i^{th}$ element of the $k^{th}$ column of the OFs $P$ matrix. Thus deterministic models can be thought of as reproducing $D$ by convexly summing over a set of $\Omega$ points in a $dm$ dimensional space. But at least $dm+1$ such points must be convexly summed for us to be able to represent an arbitrary point from $\left[0,1\right]^{dm}$. Thus we require at least $dm+1$ columns of $M$, i.e. at least $dm+1$ ontic states.



We can also set a trivial lower bound on $\Omega$:
\begin{proposition}\label{prop3}
Any ontological factorization must be such that $\Omega\ge \mathrm{rank}(D)$.
\end{proposition}
This follows immediately from the fact that $D=MP$, the ranks of
$M,P$ cannot exceed $\Omega$, and for general matrices
$\mathrm{rank}(AB)\le
\mathrm{min}(\mathrm{rank}(A),\mathrm{rank}(B))$.

As an example of where this bound can be attained in a deterministic OF, consider the data table:
\begin{equation}\label{toyD}
D= \left[
\begin{array}{cccccc}
1 & 0 & 1/2 & 1/2
& 1/2 & 1/2 \\
0 & 1 & 1/2 & 1/2 & 1/2 & 1/2 \\\hline
1/2 & 1/2 & 1 & 0 & 1/2 & 1/2 \\
1/2 & 1/2 & 0 & 1 & 1/2 & 1/2 \\\hline
1/2 & 1/2 & 1/2 & 1/2 & 1 & 0 \\
1/2 & 1/2 & 1/2 & 1/2 & 0 & 1 \\
\end{array}
\right].
\end{equation}
This is the data table for a qubit, where the measurements are the
projectors onto the eigenbases of the Pauli $X,Y,Z$ matrices, and
the states are simply the eigenstates of these three operators. This
data table has the following OF:
\begin{equation}\label{toyMP}
M=\left(
\begin{array}{cccc}
1 & 1 & 0 & 0 \\
0 & 0 & 1 & 1 \\\hline
1 & 0 & 1 & 0 \\
0 & 1 & 0 & 1 \\\hline
1 & 0 & 0 & 1 \\
0 & 1 & 1 & 0 \\
\end{array}
\right),
P=\frac{1}{2}
\left(
\begin{array}{cccccc}
1 & 0 & 1 & 0 & 1 & 0 \\
1 & 0 & 0 & 1 & 0 & 1 \\
0 & 1 & 1 & 0 & 0 & 1 \\
0 & 1 & 0 & 1 & 1 & 0 \\
\end{array}
\right).
\end{equation}

This OF has $\Omega=4$, which is less than both $s=6$ or $d^m=8$,
and which happens to be the rank of $D$. (Note that this OF is
equivalent to a single toy-bit of Spekkens \cite{toy_theory}.)

\subsection{Ontological compression: Method 1 \label{SEC:ont_comp1}}

The first technique for ontological compression we consider can be
applied to any OF, but finds particular utility in compressing the
deterministic OF's generated from Model 1 (discussed in Section
\ref{DetFromIndet}). The key is to consider the possibility that two
columns $j$ and $j^{\prime}$ of $M$, corresponding to ontic states
associated with distinct and non-orthogonal quantum states, are
identical. If this is the case, then one of the ontic states is
redundant, since any row of $P$ (i.e. any quantum state) which has
support at $j$, could just as well reproduce the relevant statistics
in $D$ by instead having support at $j^{\prime}$.

For example, consider the data table given in (\ref{toyD}) and
suppose that we started with a deterministic un-compressed OF of the
kind discussed in section \ref{DetFromIndet} so that,

\begin{equation}
M=\left(
\begin{array}{|cc|cc|cc|cc|cc|cc|}
\hline
1 & 1 & 0 & 0 & 1 & 0 & 1 & 0 & 1 & 0 & 1 & 0\\
0 & 0 & 1 & 1 & 0 & 1 & 0 & 1 & 0 & 1 & 0 & 1\\\hline
1 & 0 & 1 & 0 & 1 & 1 & 0 & 0 & 1 & 0 & 0 & 1\\
0 & 1 & 0 & 1 & 0 & 0 & 1 & 1 & 0 & 1 & 1 & 0\\\hline
{\color{Blue}0} & {\color{Blue}1} & 0 & 1 & {\color{Red}0} & {\color{Red}1} & {\color{Green}1} & {\color{Green}0} & 1 & 1 & 0 & 0\\
{\color{Blue}1} & {\color{Blue}0} & 1 & 0 & {\color{Red}1} & {\color{Red}0} & {\color{Green}0} & {\color{Green}1} & 0 & 0 & 1 & 1\\
\hline
\end{array}\right), \label{toyMbig}
\end{equation}
\begin{equation}
P=\frac{1}{2} \left(
\begin{array}{cccccc}
1 & 0 & 0 & 0 & 0 & 0 \\
1 & 0 & 0 & 0 & 0 & 0 \\
0 & 1 & 0 & 0 & 0 & 0 \\
0 & 1 & 0 & 0 & 0 & 0 \\
0 & 0 & 1 & 0 & 0 & 0 \\
0 & 0 & 1 & 0 & 0 & 0 \\
0 & 0 & 0 & 1 & 0 & 0 \\
0 & 0 & 0 & 1 & 0 & 0 \\
0 & 0 & 0 & 0 & 1 & 0 \\
0 & 0 & 0 & 0 & 1 & 0 \\
0 & 0 & 0 & 0 & 0 & 1 \\
0 & 0 & 0 & 0 & 0 & 1 \\
\label{toyPbig}
\end{array}
\right).
\end{equation}

We can immediately see that several of the columns of
(\ref{toyMbig}) are already equal (specifically, columns $1$ and
$5$, $2$ and $7$, $3$ and $12$ and $4$ and $10$). We can therefore
compress the pairs of associated ontic states together. In general
however we may not be so fortunate, and we will need to manipulate
the columns of $M$ to try and make as many identical as possible. So
long as we respect the constraints imposed by stochasticity and the
requirement $MP=D$ we are free to alter the entries of $M$ as is
convenient. In particular, we can change which ontic states a given
indicator function (i.e. row of $M$) uses to reproduce any quantum
states statistics within $D$ so long as we respect the requirement
that one and only one outcome of each PVM should ever occur.
Consider the $2\times{2}$ boxes we have drawn on $M$ in
(\ref{toyMbig}). These partition rows corresponding to distinct PVM
measurements and sets of ontic states (columns) associated with
different preparations \footnote{These sets of ontic states are
guaranteed to be disjoint since Model 1 is $\psi$-ontic.}. In terms
of these boxes, the allowed manipulations of $M$ correspond to
flipping the columns of a box,

\[
\begin{array}{|cc|}\hline
a\textrm{ } & b \\
c\textrm{ } & d
\\\hline
\end{array}\;\rightarrow
\begin{array}{|cc|}\hline
b\textrm{ } & a \\
d\textrm{ } & c
\\\hline
\end{array}\:{.}\
\]

As an example of how one can use this freedom to compress two ontic
states, columns $3$ and $6$ of (\ref{toyMbig}) can be made equal by
simply applying this permutation to the bottom box containing
columns $5$ and $6$ (shown in {\color{Red}red}). There is however, a
subtlety in forcing columns of $M$ to be equal in this way. The
permutation we apply to make columns $3$ and $6$ equal will of
course also alter column $5$. But initially this column was already
equal to column $1$. To retain this initial equality we must apply
the same permutation to the bottom box containing columns $1$ and
$2$ (shown in {\color{Blue}blue}). However, this in turn will alter
column $2$, which was initially identical to column $7$ and so to
retain \textit{this} equality we must then permute the bottom box
containing columns $7$ and $8$ (shown in {\color{Green}green}).
Performing these three permutations we obtain the following new
expression for $M$,

\begin{equation}
M=\left(
\begin{array}{cccccccccccc}
1 & 1 & 0 & 0 & 1 & 0 & 1 & 0 & 1 & 0 & 1 & 0\\
0 & 0 & 1 & 1 & 0 & 1 & 0 & 1 & 0 & 1 & 0 & 1\\\hline
1 & 0 & 1 & 0 & 1 & 1 & 0 & 0 & 1 & 0 & 0 & 1\\
0 & 1 & 0 & 1 & 0 & 0 & 1 & 1 & 0 & 1 & 1 & 0\\\hline
1 & 0 & 0 & 1 & 1 & 0 & 0 & 1 & 1 & 1 & 0 & 0\\
0 & 1 & 1 & 0 & 0 & 1 & 1 & 0 & 0 & 0 & 1 & 1\\
\end{array}\right). \label{toyMequals}
\end{equation}

\begin{figure*}[t]
\includegraphics[scale=0.8]{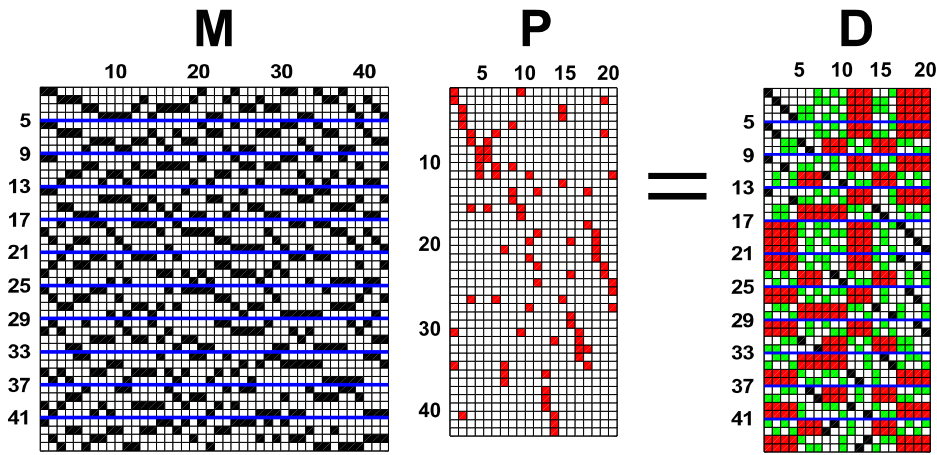}\caption{(Color online) Compressed P and M matrix for a deterministic OF of the Kernaghan set of
preparations and measurements. Color coding as in Figure
\ref{FIG:peresPM}. Note that there is now some overlap in the ontic
states associated with different preparations (cf. the different
columns of $P$).} \label{FIG:peresPMcompress}
\end{figure*}

Combined with (\ref{toyPbig}) this reproduces the statistics from
$D$. Counting identical columns we find that in fact there are now
$8$ redundant columns in $M$, three more having appeared due to the
extra reshuffling we needed to perform in order to share columns $3$
and $6$. These redundant columns can be removed, which results
in us compressing the four triples $(1,5,9)$, $(2,7,11)$,
$(3,6,12)$, $(4,8,10)$ into four ontic states. We can then finally
update $P$ to represent this compression by combining its rows
corresponding to each of the four triples of ontic states. The
resultant compressed OF is in fact precisely the optimally
compressed model given in (\ref{toyMP}).

In cases with larger data tables, the compression is not quite so
trivial. The Kernaghan OF for example, requires one to keep track of
eighty ontic states. By applying an algorithm for randomly combining
ontic states according to a natural generalization of the example
given above, we are able to compress the Kernaghan OF to use only
$42$ ontic states. This is much lower than the $d^m$ ontic states
needed by Model 2 and almost half the number required by our deterministic rendering of Model 1. The resulting $M$ and $P$ matrices are shown in Figure \ref{FIG:peresPMcompress}. Clearly the
columns of the matrix $P$ associated with this compressed OF are not
disjoint, and consequently the OF is $\psi$-epistemic. In fact, the
approach to compression outlined above will always result in a
$\psi$-epistemic model. It should be noted however, that the
compression illustrated in Figure \ref{FIG:peresPMcompress} is not
guaranteed to be optimal, since it does not saturate the bound given
in Proposition \ref{prop3}. In general, one would expect the amount
by which one can compress an OF using this approach to depend on the
initially chosen $M$ (picked at random for the above examples) and
the order in which one chooses to compress the various ontic states
of the model. Optimizing these choices is highly non-trivial.

\subsection{Ontological compression: Method 2}

The second method for ontological compression we consider is
designed to effect a compression for OF's based on Model 2. It is
also most easily illustrated by an example. Let us return to the
example of Eqs. (\ref{model2D}) and (\ref{model2MP}), reproduced
here:
\begin{equation*}
\!\begin{array}{c} \left[
\begin{array}{cc}
0 & 2/3 \\
1/3 & 1/3 \\
2/3 & 0 \\\hline
1/3 & 1/2 \\
1/3 & 1/2 \\
1/3 & 0 \\
\end{array}\right]\\
\\
\\
\\
\end{array}
\begin{array}{c}
\!=\!\\
\\
\\
\\
\end{array}
\begin{array}{c}\left(
\begin{array}{ccccccccc}
1 & 1 & 1 & 0 & 0 & 0 & 0 & 0 & 0 \\
0 & 0 & 0 & 1 & 1 & 1 & 0 & 0 & 0 \\
0 & 0 & 0 & 0 & 0 & 0 & 1 & 1 & 1 \\\hline
1 & 0 & 0 & 1 & 0 & 0 & 1 & 0 & 0 \\
0 & 1 & 0& 0 & 1 & 0 & 0 & 1 & 0 \\
0 & 0 & 1 & 0 & 0 & 1 & 0 & 0 & 1 \\
\end{array}
\right)\\
\\
\\
\\
\end{array}\!\!\!\!\left(
\begin{array}{cc}
0 & 1/3 \\
0 & 1/3 \\
0 & 0 \\
1/9 & 1/6 \\
1/9 & 1/6 \\
1/9 & 0 \\
2/9 & 0 \\
2/9 & 0 \\
2/9 & 0 \\
\end{array}
\right).
\end{equation*}
It is convenient to view the 9 entries corresponding to each of the
two columns of $P$ as probability distributions over 9 ``ontic state
boxes'' as follows:
\[
\mathcal{P}^{(1)}=
\begin{array}{|c|c|c|}\hline
\textrm{ }0\textrm{ } & 1/9 & 2/9 \\\hline
0 & 1/9 & 2/9 \\\hline
0 & 1/9 & 2/9 \\\hline
\end{array}\:{,}\;\;
\mathcal{P}^{(2)}=
\begin{array}{|c|c|c|}\hline
1/3 & 1/6 & \textrm{ }0\textrm{ } \\\hline
1/3 & 1/6 & 0 \\\hline
0 & 0 &  0 \\\hline
\end{array}\:{.}
\]
The indicator functions are all of the form
\[
\begin{array}{|c|c|c|}\hline
\textrm{ }1\textrm{ } & \textrm{ }0\textrm{ } & \textrm{ }0\textrm{ } \\\hline
1 & 0 & 0 \\\hline
1 & 0 & 0 \\\hline
\end{array}\:{,}\;\;\;
\begin{array}{|c|c|c|}\hline
\textrm{ }1\textrm{ } & \textrm{ }1\textrm{ } & \textrm{ }1\textrm{ } \\\hline
0 & 0 & 0 \\\hline
0 & 0 &  0 \\\hline
\end{array}\:{,}\:\textrm{ etc. }
\]

\noindent That is, the construction of Model 2 is such that summing
$\mathcal{P}^{(k)}$ along a column of the boxes yields the
probability for the corresponding outcome of the first measurement,
summing along a row yields the same for the second measurement. Thus
it is clear that if we leave the row and column box-sums invariant
we can ``shift around'' the probability weightings that the
distributions $\mathcal{P}^{(1)},\mathcal{P}^{(2)}$ assign to the
ontic states. If we then find that both distributions assign a
weighting 0 to an ontic state, it can be removed completely and some
ontological compression has been achieved. In fact,  we already see
that $\mathcal{P}^{(1)}_{3,1}=\mathcal{P}^{(2)}_{3,1}=0$, and so
this ontic state could be removed.

In order to shift around the probability, one simple operation
involving 4 ontic states is to map them as follows:
\begin{eqnarray*}
\mathcal{P}_{i,j}&\rightarrow& 0,\\
\mathcal{P}_{i,j'}&\rightarrow& \mathcal{P}_{i,j'}+\mathcal{P}_{i,j},\\
\mathcal{P}_{i',j}&\rightarrow& \mathcal{P}_{i',j}+\mathcal{P}_{i,j},\\
\mathcal{P}_{i',j'}&\rightarrow&
\mathcal{P}_{i',j'}-\mathcal{P}_{i,j}.
\end{eqnarray*}

Since the final state must be a suitable probability distribution,
we require $\mathcal{P}_{i',j'}>\mathcal{P}_{i,j}$. In our example
we can use this to implement:
\[
\begin{array}{lcl}
\mathcal{P}^{(1)}_{2,2}\rightarrow 0&   \textrm{ }& \mathcal{P}^{(2)}_{2,2}\rightarrow 0,\\
\mathcal{P}^{(1)}_{2,3}\rightarrow 3/9&\textrm{ } &\mathcal{P}^{(2)}_{2,1}\rightarrow 1/2, \\
\mathcal{P}^{(1)}_{3,2}\rightarrow 3/9&\textrm{ } &\mathcal{P}^{(2)}_{1,2}\rightarrow 1/3, \\
\mathcal{P}^{(1)}_{3,3}\rightarrow 1/9&\textrm{ } &\mathcal{P}^{(2)}_{1,1}\rightarrow 1/6. \\
\end{array}
\]

In this manner we nullify weighting of both distributions on the
middle ontic state (i.e. $\mathcal{P}^{(k)}_{2,2}$=0) , and it too
could be removed:

\[
\mathcal{P}^{(1)}=
\begin{array}{|c|c|c|}\hline
\textrm{ }0\textrm{ } & 1/9 & 2/9 \\\hline 0 & \blacksquare & 3/9
\\\hline \blacksquare & 3/9 & 1/9 \\\hline
\end{array}\:{,}\;\;
\mathcal{P}^{(2)}=
\begin{array}{|c|c|c|}\hline
1/6 & 1/3 & \textrm{ }0\textrm{ } \\\hline 1/2 & \blacksquare & 0
\\\hline \blacksquare & 0 &  0 \\\hline
\end{array}\:{.}
\]

(Where $\blacksquare$ denotes entries to be deleted.)

\subsection{Classical simulation and ontological compression \label{SEC:class_simulation}}

We have seen that some sort of ontological compression may be
possible, depending on specifics of the data tables involved. Why
might one be interested in ontological compression? Consider a
``family'' of data tables - that is a set of data tables constructed
from an increasing number of states and measurements. An example
might be  data tables constructed from the family of all stabilizer
states on $n$-qubits, for increasing values of $n$. If it were the
case that this data table can be ontologically compressed such that
$\Omega$ only grows polynomially with $n$, then an efficient
classical simulation of that family of states and measurements is
clearly possible.   If we denote by $D_{n}$ the data table for all $n$-qubit stabilizer
states/measurements so that $D_{n}$ has $O\left( 4^{n}\right) $ rows/columns
(and $D_{1}$ would then be the data table of (\ref{toyD})), then we are asking
whether there exists an ontological factorization of $D_{n}$ which has $%
\Omega =O\left(\text{poly}\left(n\right)\right)$. But we can immediately
see that this cannot be achieved, because any table containing all
stabilizer states and measurements will contain a sub-table in which (for $d=2$) $2^{n}$
preparations are associated with all $n$ bit strings over $n$ measurements -
i.e. a sub-table of the form shown in (\ref{table_upper_bound}). Since
we have already seen that such a table will require $\Omega \geq 2^{n}$,
then its existence clearly prohibits us from finding ontological factorization of
the stabilizer data table having $\Omega =O\left(\text{poly}\left(n\right)\right)$.

Actually, a polynomial sized number of ontic states is too strong a
requirement for efficiency: Classical Monte-Carlo simulations, for
example, can be provably efficient for certain problems even when
the number of ontic states grows exponentially. Since our
ontological models are also essentially classical probability
representations, we similarly expect a polynomial requirement to be
too strong for them.

\subsection{Connections to other matrix factorization problems \label{SEC:matrix_connection}}

We conclude this section with a brief discussion of how finding OF's
is related to other matrix factorization problems. Our data tables
are generally not square. However, in instances where (such as in
the example of Eqs. (\ref{toyD}) and (\ref{toyMP})) we use the same
quantum states to specify both the preparation procedures and the
measurements, then $D$ is symmetric, with non-negative entries, and
positive semi-definite (ie doubly non-negative). These conditions
are necessary (though not sufficient) for the matrix $D$ to be
completely positive - that is,  factorizable into a product of the
form $A=B^\dagger B$, where the elements of $B$ are non-negative.
Completely positive matrices have been much studied (see e.g.
\cite{cpmatrices}), in particular with regards to finding (bounds
on)  the smallest row-dimension of $B$ for which this factorization
is possible (called the ``cp-rank'' of $B$, or sometimes the
``factorization index'').

A completely positive matrix factorization (i.e. one of the form
$D=B^\dagger B$) is not generally equivalent to our OF's. However,
we have already seen an example of such a factorization appearing as
an OF  - the matrices $M,P$ of the example in Eqs. (\ref{toyD}) and
(\ref{toyMP}) differ only by an overall multiplicative factor of
$1/2$, and so we can see this example as providing a
completely-positive matrix factorization of this particular $D$. Now
it is interesting to note that there is an upper bound \footnote{Note
that if we are interested in data tables constructed from quantum
states $|\psi_i\>$ and rank 1 operators $|\phi_j\>\<\phi_j|$ in a
fixed Hilbert space dimension $d_Q$, then finding such non-trivial
upper bounds on $\Omega$ cannot be achieved by simply finding a
family of data tables of continually increasing rank and using
Proposition \ref{prop3}. This is because $\mathrm{rank}(D)\le
d_Q^2$. (To see this, note that $D=C\circ C^*$ where
$C_{ij}=\<\phi_i|\psi_j\>$ and $\circ$ denotes the Hadamard
(elementwise) product - the result follows from standard results
regarding this product \cite{hornjohnson} and from the fact that
$\mathrm{rank}(C)\le d_Q$.)} on the cp-rank \cite{barioli} which is
polynomial in $\mathrm{rank}(D)$, which, in turn, is polynomial in
$d_Q$. The cp-rank is, in the OF picture, the number of ontic states
$\Omega$. Thus it seems plausible there exist some interesting data
table families which have polynomial (in $d_Q$) sized OFs.

There is a second type of matrix factorization which has some
connection to the OFs we have been considering: the so-called
``non-negative matrix factorization'' (NMF). In this factorization a
matrix of (non-negative-valued) image data $V$ is factorized into
the product of two matrices $W,H$ with non-negative elements:
\[ V=WH. \]

NMF is often considered from the viewpoint of image compression,
where the goal is to obtain an approximate factorization where $W$
($H$) have as small a number of rows(columns) as possible. Some
particularly simple procedures for performing this approximate
factorization, guaranteed to converge to a local optima (with
respect to various matrix distances between the ideal $V$ and the
approximate one), were given in \cite{nmf}. It was shown recently in
\cite{ho} that if one considers optimizing with respect to the
Kullback-Leibler distance, then the local optima preserve the row
and column sums of the original matrix. In particular, if we
consider the separate NMF factorizations \[ D^{(x)}=M^{(x)}P,\] so
$D^{(x)}$ is column-stochastic, then the results of Corollary 2 in
\cite{ho} imply that the NMF found will be such that $M^{(x)},P$ are
column stochastic as we require. Of course our problem requires
finding many such factorizations with the \emph{same} $P$ for the
$m$ different $D^{(x)}$'s. Interestingly, one reason that NMF is
useful in image analysis is that it breaks the image up into
``hidden variables'' (Lee and Seung's terminology, not ours!) such
that an image is comprised of pieces which humans recognise as
familiar fundamental components (the ears, nose and mouth of a
facial image for example). This is unlike the more standard
principal component analysis route to ``eigenimage'' compression. In
these terms the OF we are considering is asking an interesting
question regarding factorizing many different images (the
$D^{(x)}$'s) using the same set of fundamental component images (the
$P$).
\\

\section{Conclusions}
By using an ontological formalism to describe finite sets of
preparations and measurements performed on a quantum system we have
illustrated how many features of ontological models are manifested
in such discretized scenarios. We have shown how contextuality and
deficiency appear in this formalism and have also been able to use
the formalism to build a $\psi$-epistemic model of a discrete set of
quantum statistics. We have also seen how any indeterministic
description of a finite data table can be developed into an equally
valid deterministic one and noted that the same technique (first
employed by Bell) can be applied in the continuum limit. We have
also discussed how OFs can be used to compress the number of ontic
states needed to describe such discretized quantum scenarios and
speculated on repercussions which this might have for classical
simulations of quantum systems.

\end{document}